\documentclass[10pt,letterpaper,twocolumn,aps,showpacs,preprintnumbers,amsmath,amssymb,nofootinbib,prl]{revtex4-2}
\usepackage{graphicx}% Include figure files
\usepackage{dcolumn}% Align table columns on decimal point
\usepackage{bm}% bold math
\usepackage[utf8]{inputenc}
\usepackage[T1]{fontenc}

\usepackage{color}
\usepackage{amsmath}
\usepackage{amssymb}
\usepackage[unicode=true,pdfusetitle, bookmarks=true,bookmarksnumbered=false,bookmarksopen=false,
 breaklinks=false,pdfborder={0 0 0},pdfborderstyle={},backref=false,colorlinks=true]
 {hyperref}
\hypersetup{
 pdfborderstyle={},colorlinks,linkcolor=red,citecolor=blue}
\makeatletter
\usepackage{setspace}
\usepackage{mathrsfs}
\usepackage{siunitx}
\usepackage{soul}

\makeatother
\begin{document}
\title[\texorpdfstring{$\quad$}{}]{Observation of quantum noise reduction in a Raman amplifier via quantum correlation between atom and light}

\author{Jianmin Wang\texorpdfstring{$^{*}$}{}, Rong Zhu\texorpdfstring{$^{*}$}{}, Yue Li\texorpdfstring{}{},  Z. Y. Ou\texorpdfstring{$^{\dag}$}{}} 

\affiliation{\mbox{Department of Physics, City University of Hong Kong, 83 Tat Chee Avenue, Kowloon, Hong Kong}\\ 
\texorpdfstring{$^{*}$}{}equal contribution\\
\texorpdfstring{$^{\dag}$}{}jeffou@cityu.edu.hk}

\date{\today}

\begin{abstract}
Any amplifier requires coupling to its internal degrees of freedom for energy gain. This coupling introduces extra quantum noise to the output. On the other hand, if the internal degree of the amplifier can be accessed and manipulated, we can manage and even reduce the quantum noise of the amplifier's output.  In this paper, we present an experiment to reduce the quantum noise of a Raman amplifier by preparing the atomic medium in a correlated state with the Stokes light field. We report an observation of quantum noise reduction of more than 3.5 dB in the atomic Raman amplification process. From another perspective, the Raman amplifier at high gain in turn serves as a measurement tool for the quantum correlation between the atom and light. Furthermore, such a scheme, when viewed as a whole, also forms a quantum-entangled atom-light hybrid interferometer that can lead to quantum-enhanced sensors.
\end{abstract}

\maketitle

It is known that extra noise is added to the output of an amplifier in addition to the amplified input noise even for an ideal amplifier with all classical noise eliminated, leading to degradation of the signal-to-noise ratio \cite{haus}. The source of this excess noise is from the quantum fluctuations of the amplifier's internal degrees, which are usually unattended and in vacuum state. This excess noise becomes important at the quantum level \cite{hong85,gla86}. Caves was the first to formulate a general theory for a quantum amplifier to incorporate internal degrees \cite{cav82}. On the other hand, if the internal degree can be accessed and manipulated, excess noise can be managed. So, based on the model by Caves, it was proposed by Milburn et al. \cite{mil87} and demonstrated experimentally by Ou et al. \cite{ou93} that placing the internal degrees in squeezed states can reduce such excess noise, leading to noiseless amplification. Furthermore, since quantum entanglement can also give rise to noise reduction in joint quantities \cite{ou92PRL}, it was proposed by Ou \cite{ou93b} and experimentally demonstrated by Kong et al.~\cite{kong13} that when the internal degree is prepared in a quantum correlated state with the input field of the amplifier, the reduction of quantum noise of amplification can be achieved. 

\begin{figure}[ht]
\centering
\includegraphics[width=1.0\columnwidth]{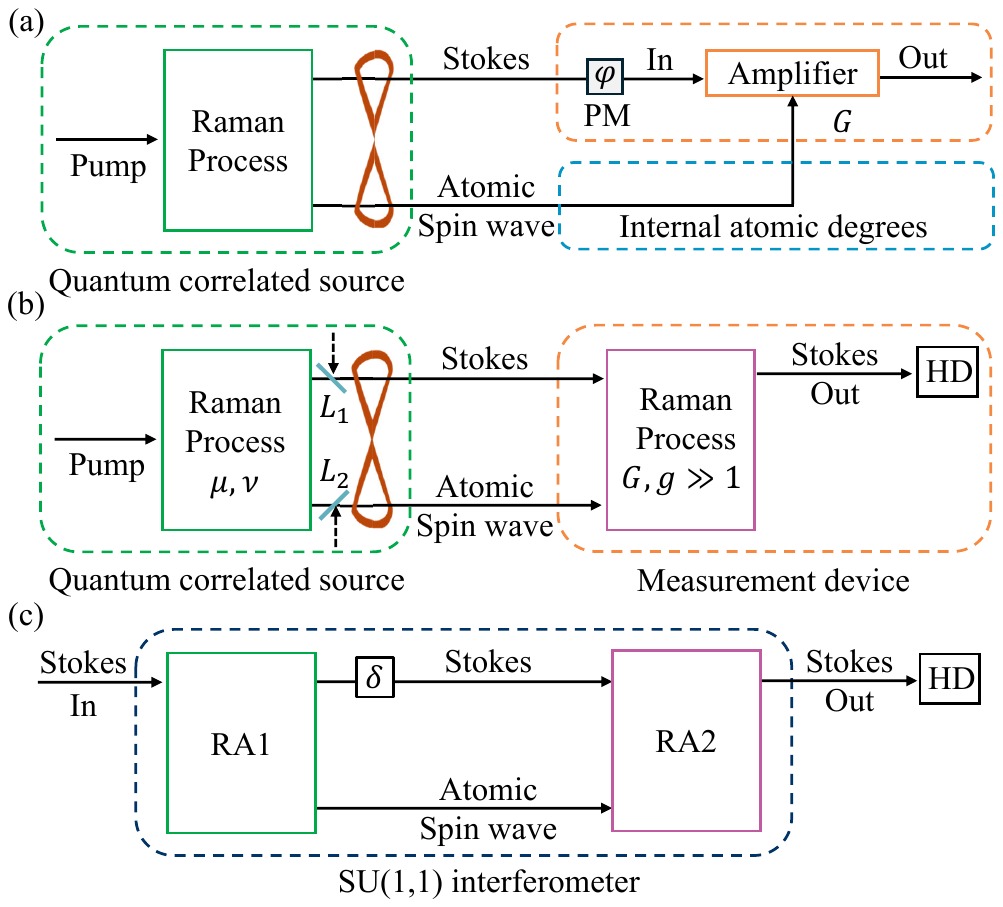}
	\caption{Conceptual schemes for (a) quantum noise reduction of an amplifier by correlating internal degree with the input field; (b) measurement of quantum correlation/entanglement between atomic spin wave and Stokes light field by a Raman amplifier; (c) formation of a hybrid atom-light SU(1,1) interferometer using Raman amplifiers (RA) as beam splitters for the input Stokes field.} 
	\label{Fig1}
\end{figure}

So far, the theory and experiment were mostly done on optical parametric amplifiers where the internal degree is the idler optical mode of the amplifier, which can be easily coupled optically from outside. However, for an amplifier using atoms as the gain medium,  internal atomic degrees are usually not directly accessible, although the atomic states have been prepared in an atomic spin-squeezed state through optical interaction \cite{bao20}. So, although amplifiers based on active atomic media are widely used in lasers and optical communications and their noise performance has been well known, there was no effort for the reduction of their quantum noise because most of the applications are well beyond the quantum noise limit. Furthermore, recent work on atom-light hybrid SU(1,1) interferometers \cite{chen15} employed Raman amplifiers as beam splitters, which are supposed to operate at the quantum level \cite{ou20}, but no studies were carried out.  In this paper, we study in detail the quantum noise performance of a Raman amplifier based on the atomic medium of Rb-87 in a hot vapor cell.  As shown in Fig.\ref{Fig1}(a), we first prepare the atomic state in quantum correlation with the input Stokes field through another Raman process. Then we use the prepared atoms for Raman amplification of the correlated Stokes input. We measure the quantum noise level of the output field of the Raman amplifier under various conditions and observe a reduction in noise by more than 3.5 dB compared to the case when atoms are in an unattended state. 

Recently, it was discovered that a high-gain parametric amplifier can be used to measure the quantum noise reduction and quantum correlation between different fields \cite{shaked18,li19}. This technique can overcome large losses in propagation before detection \cite{li19,li20} and was used to measure on-chip quantum squeezing even in the presence of huge coupling losses \cite{neh22}. So, from the perspective of the atomic medium and the Stokes field generated in the first Raman amplifier, the second Raman amplifier can serve as a tool for the measurement of their quantum correlation when it operates at high gain \cite{shaked18,li19,li20,bel21}, as shown in Fig.\ref{Fig1}(b). Our analysis shows a quantum correlation coefficient of nearly -4 dB below the uncorrelated case, demonstrating the quantum entanglement between the atom and the Stokes light. This is the first study of the quantum noise correlation between atoms and light in a Raman amplifier. Furthermore, the geometry of two Raman amplifiers (RA) can be viewed as an SU(1,1) interferometer with each RA serving as beam splitter and combiner for the input Stokes field, as shown in Fig.\ref{Fig1}(c), which can be used as a quantum sensor through change of atomic state.

The conceptual schematic of an optical amplifier with atoms as the gain medium is shown in Fig.\ref{Fig1}(a), where the internal atomic degrees of the amplifier are coupled to the input and add extra noise to the output. In general, the input and output field operators of the amplifier are related by \cite{cav82} 
\begin{equation}\label{amp}
    \hat a_{out} = G\hat a_{in} + \hat F
\end{equation}
where $G$ is the amplitude gain parameter and $\hat F$ describes the internal degrees of the amplifier that normally add extra noise to the amplified output. 

In particular, the quantum model for a Raman amplifier was first formulated by Raymer and Mostowski \cite{ray81}, which follows the general expression of Eq.(\ref{amp}). For an atomic Raman amplifier pumped by a strong classical field $A_P$, the signal field is the Stokes field $\hat a_S$ and they are coupled through a group of $N_a$ atoms via two lower states $|{\rm g}\rangle, |{\rm m}\rangle$ by an effective Raman interaction Hamiltonian \cite{polzik10}:
\begin{eqnarray}\label{Hr}
\hat H_R= i\hbar \eta A_{P} \hat a_{S}^{\dagger}\hat S_A^{\dagger}
-i\hbar \eta \hat a_S\hat S_AA_{P}^{*},
\end{eqnarray}
where 
\begin{equation}
    \hat S_A = \frac{1}{\sqrt{N_a}}\sum_i |{\rm g}\rangle_i\langle {\rm m}| ,
\end{equation}
is the field operator for collective atomic excitation, also known as the atomic (pseudo) spin wave, that connects the ground state $|{\rm g}\rangle$ and the meta-stable state $|{\rm m}\rangle$ and couples to the Stokes light field $\hat a_S$ and the Raman pump field via an upper excited level $|{\rm e}\rangle$. The interaction coefficient $\eta$ is given by $\eta = g_{\rm eg}g_{\rm em}/\Delta$ with $g_{\rm eg},g_{\rm em}$ as the coupling coefficients between the excited state and the lower-level states and $\Delta$ as the detuning from the excited state for both Stokes field $\hat a_S$ and Raman pump field $A_P$. 

After some interaction time $t$, the Raman interaction Hamiltonian in Eq.(\ref{Hr}) gives rise to input-output relations as
\begin{eqnarray}\label{in-out}
\hat a_{S}^{(out)} &=& \hat a_S^{(in)} \cosh |\eta A_P|t + \hat S_A^{(in) \dagger} \sinh |\eta A_P|t\cr
\hat S_A^{(out)} &=& S_A^{(in)} \cosh |\eta A_P|t + \hat a_S^{(in) \dagger} \sinh |\eta A_P|t.
\end{eqnarray}
Comparing to Eq.(\ref{amp}), we have $G=\cosh |\eta A_P|t$, and $\hat F= \hat S_A^{\dagger}\sinh |\eta A_P|t$ which is directly related to the internal atomic states of the amplifier. 

Normally, the atoms are in the ground state $|{\rm g}\rangle$ and independent of the input Stokes field. Then, the Raman interaction makes them transition to the meta-stable state $|{\rm m}\rangle$ and in the meantime a Stokes photon emits spontaneously, which adds extra noise to the amplified Stokes signal. This can be seen from the output noise in the quadrature-phase amplitude $\hat X_S \equiv \hat a_S +\hat a_S^{\dagger}$:
\begin{eqnarray}\label{noise}
\langle \Delta^2 \hat X_S^{(out)}\rangle &= &G^2 \langle \Delta^2 \hat X_S^{(in)}\rangle + \langle \Delta^2 \hat X_A^{(in)}\rangle (G^2-1)\cr
&=& G^2 \langle \Delta^2 \hat X_S^{(in)}\rangle + (G^2-1)\cr
&=& 2G^2-1 ~~~~{\rm for~ coherent~state~input}.
\end{eqnarray}
Here, $\hat X_A^{(in)} \equiv \hat S_A^{(in)}+\hat S_A^{(in) \dagger}$ and $G_q\equiv 2G^2-1$ is defined as the quantum noise gain when all input fields are in vacuum, where $\hat X_A$ and $\hat X_S$ represent the atomic spin wave and Stokes light field, respectively.
% However, if the internal atomic states are prepared in a correlated state with the input field, that is, atomic spin wave $\hat S_a^{(in)}$ is related to the input Stokes field $\hat a_S^{(in)}$, we have from Eq.(\ref{in-out})
However, we notice Eq.(\ref{in-out}) can be rewritten in terms of quadrature-phase amplitudes as $\hat X_S^{(out)} =G \hat X_S^{(in)} + \hat X_A^{(in)}\sqrt{G^2-1}$,  from which we obtain:
\begin{eqnarray}\label{noise2}
\langle \Delta^2 \hat X_S^{(out)}\rangle_{qc} &= &G^2 \langle \Delta^2( \hat X_S^{(in)} + \lambda \hat X_A^{(in)})\rangle 
\end{eqnarray}
with $\lambda \equiv \sqrt{G^2-1}/G \rightarrow 1 (G\gg 1)$. So, output noise of the amplifier can be reduced if the internal atomic states are prepared in a correlated state with the input field, that is, the atomic spin wave $\hat X_A^{(in)}$ is anti-correlated with the input Stokes field $\hat X_S^{(in)}$. Here, the subscript $``qc"$ denotes the correlated input.  More specifically, comparing Eq.(\ref{noise}) and Eq.(\ref{noise2}), we have 
\begin{eqnarray}\label{R}
R\equiv \frac{\langle \Delta^2 \hat X_S^{(out)}\rangle_{qc}}{\langle \Delta^2 \hat X_S^{(out)}\rangle} &= & \frac{\langle \Delta^2( \hat X_S^{(in)} + \lambda \hat X_A^{(in)})\rangle}{1+\lambda^2}.
\end{eqnarray}
Noise reduction $R<1$ requires quantum correlation to reduce $\langle \Delta^2( \hat X_S^{(in)} + \lambda \hat X_A^{(in)})\rangle$ below the uncorrelated value of $1+\lambda^2$.

Normally, it is not easy to create quantum correlations between the light field and atoms. However, quantum correlations are known to be generated in a parametric amplifier between its signal and idler fields \cite{ou92}. Looking at Eqs.(\ref{Hr},\ref{in-out}), we find the Raman amplifier described there indeed corresponds to a parametric amplifier \cite{shaked18,li20} but with its idler field replaced by the atomic spin wave $\hat S_A$. So we can create a quantum correlation between the atoms and the Stokes field by another Raman process. This correlation between atoms and the Stokes field from a Raman process was first confirmed in photon/atom numbers \cite{kuz03}, but the correlation in quadrature-phase amplitudes, which has not been observed so far, is what we will study here and exploit for quantum noise reduction.
\begin{figure}[t]
\centering
\includegraphics[width=1.0\columnwidth]{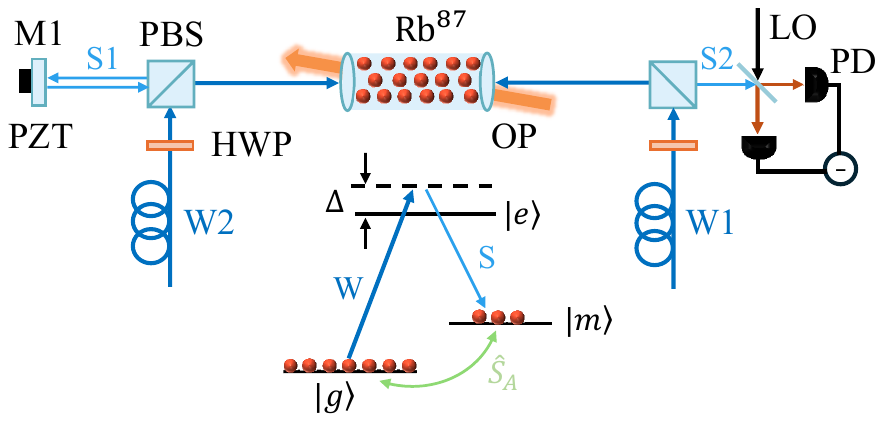}
	\caption{Experimental schematic for a Raman amplifier (RA2) with the atoms and light field prepared in a quantum correlated state by another Raman process (RA1). M: mirror; PBS: polarization beam splitter; PZT: piezo-electric transducer. HWP: half wave plate. LO: local oscillator. S: Stokes field. PD: photo diode. OP: Optical pumping.} 
	\label{Fig2}
\end{figure}

% (Describe here in detail the Raman process depicted in the inset). Figure \ref{Fig2} shows the experimental schematic for preparing the atomic medium of a Raman amplifier in quantum correlation with the input light field. 

 The experimental setup is shown in Fig~\ref{Fig2}. Pure Rb-87 is contained in a 75~mm-long glass cell. It is placed in four-layer $\mu$-magnetic shielding to reduce the stray Earth magnetic field and the cell is heated to 70°C by a heating belt. The lower two energy states $|{\rm g}\rangle$ and $|{\rm m}\rangle$ are the hyperfine split ground states $5 ^2S_{1 / 2}(F=1,2)$ with a frequency difference of 6.87~GHz and the higher energy states $|{\rm e}\rangle$ are the excited states $5 ^2P_{1 / 2}(F=2)$, where the detuning $\Delta$ of the higher energy states is around 800~MHz. The continuous light W1 and W2 in opposite directions serve as the pumping fields for the Raman amplifiers RA1 and RA2, respectively, with corresponding Stokes fields S1 and S2 also in opposite directions. Optical pumping is used for initial preparation of the atoms in the ground state. The Raman amplifier that we want to characterize its noise behavior is pumped by W2 which interacts with a group of hot Rubium-87 atoms. Its output Stokes field S2, which has a polarization opposite to that of the Raman pump field and is separated from it by a PBS, is measured by a balanced homodyne detector for the quantum noise level of the output of the amplifier. Normally, the input field to RA2 is independent of the atomic states. To create quantum correlation between the atomic states and the input field, we use the output field S1 of RA1, which is now correlated with the atomic states, as the input to RA2 by reflecting back with a mirror and re-injecting to interact with the same group of atoms. We investigate the effect of RA1 on the quantum noise level of RA2.

\begin{figure}[t]
\centering
\includegraphics[width=1.0\columnwidth]{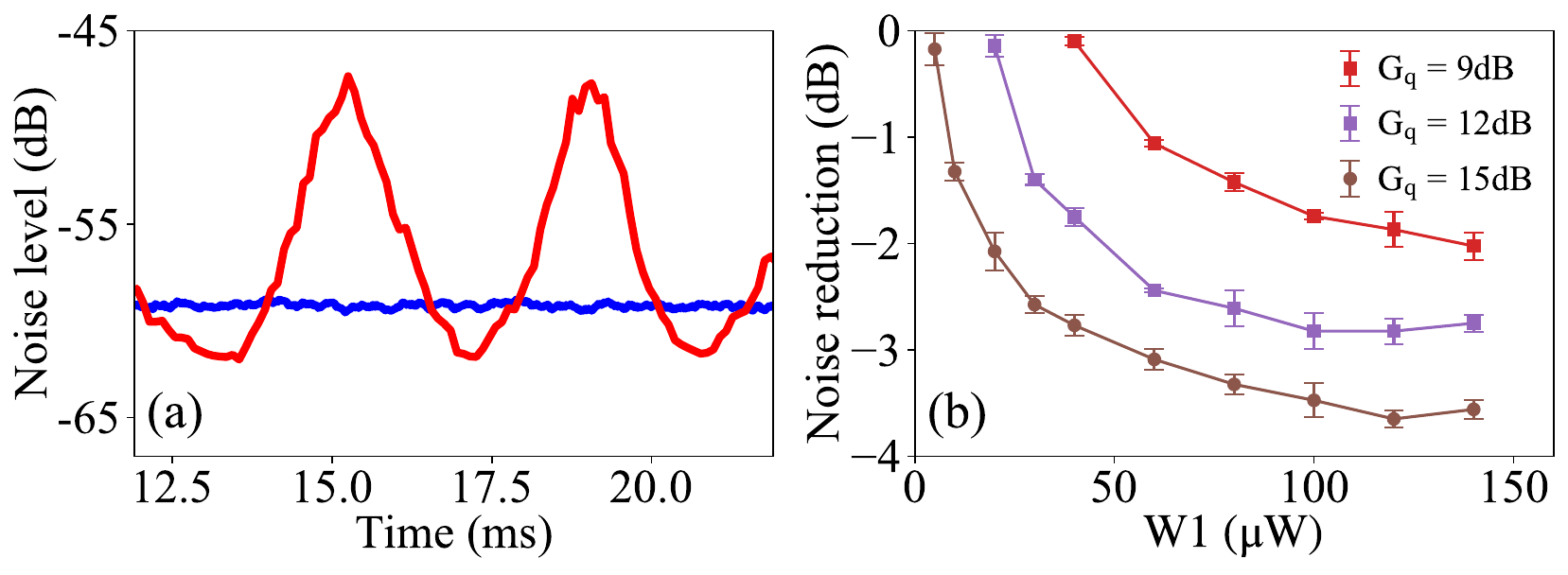}
	\caption{Quantum noise reduction of a Raman amplifier. (a) Noise levels for  uncorrelated input (blue), correlated input with a phase scan (red) and the vacuum or shot noise level is -73.6~dB; (b) Noise reduction as a function of  pump power of RA1 for various gains of RA2.} 
	\label{Fig3}
\end{figure}

As a reference of comparison, we first record the noise level of the output Stokes field S2 of RA2 without injection of S1 by blocking W1. This corresponds to the case where the atomic medium is in a state independent of the input field and the measured noise level is simply the amplified vacuum noise level. The result is shown in Fig.\ref{Fig3}(a) as the blue trace, which is about 15 dB above the vacuum noise level. This gives a quantum noise gain of $G_q=15$ dB for the Raman amplifier RA2. Then we turn on RA1 by unblocking W1 to create a quantum correlation between the atomic medium and the input field (S1) to RA2. The measured output Stokes noise level is shown in Fig.\ref{Fig3}(a) as the red trace. It turns out that the quantum correlation between the atoms and the light (S1) created by RA1 depends on the phase of the light field, its effect on RA2 is also phase-sensitive. So, we scan the phase of the reinjected Stokes field S1 with a piezoelectric transducer. The red trace of the measured noise level in Fig.\ref{Fig3}(a) goes up and down as the phase is scanned, showing this dependence. Notice that the minimum of the red trace is about 3 dB below the blue trace, indicating a quantum noise reduction of the Raman amplifier RA2.
 
Since the noise reduction effect relies on the degree of quantum correlation between the atoms and the light fields, which in turn depends on the pump power of W1, we measure the amount of noise reduction as a function of the power of W1 and plot the results in Fig.\ref{Fig3}(b) at various gains of RA2. As can be seen, the noise reduction increases with the increase of W1, showing the effect of quantum correlation. Notice in Fig.\ref{Fig3}(b) that the noise reduction amount increases with the quantum noise gains of RA2. Since the access noise of the amplifier relies on the gain of the amplifier, this shows the reduction of the access noise from the internal degree of the amplifier due to its correlation with the input.

%The quantum correlation between atoms and the Stokes field is normally characterized by the quantity $\langle \Delta^2(\hat X_a^{(in)} + \hat X_S^{(in)})\rangle \equiv \Delta^2 X_+$. The traditional way to measure this correlation is by measuring $\hat X_a^{(in)}$ and $\hat X_S^{(in)}$ separately and comparing them \cite{ou92}. But it is difficult to directly measure the atomic quantity $\hat X_S^{(in)}$. So, although it is known that there exists a quantum correlation between the Stokes field and atomic states of the Raman amplifier due to Raman interaction, it was never measured. However, we find from Eq.(\ref{R}) that when $\lambda =1$ or $G\rightarrow \infty$, $R = \frac{1}{2}\langle \Delta^2(\hat X_a^{(in)} + \hat X_S^{(in)})\rangle = \frac{1}{2} \Delta^2 X_+$. So, we can measure the quantum correlation between atom and light from the measurement of noise reduction $R$.  This is exactly the technique discussed in Refs.\cite{shaked18,li19,li20} to include RA2 as part of the measurement process for measuring the quantum correlation between two fields but here between atomic state and Stokes light field generated in RA1.  

% With power of W2 ($G_q$) fixed, noise reduction rises as power of W1 increases, which is in agreement with theory. This is attributed to the fact that $\mu^2-\nu^2=1$, as the power of W1 rises, $\mu$ and $\nu$ approach each other, causing the noise reduction to be observed.

%The dependence of the amount of noise reduction on the quantum correlation created by RA1 between the atomic medium and the light field provides us with a method to directly measure the degree of the quantum correlation.  

The quantum correlation between atoms and the Stokes field is normally characterized by the quantity $\langle \Delta^2(\hat X_A^{(in)} + \hat X_S^{(in)})\rangle \equiv \Delta^2 X_+$. Traditionally, quantum correlations between two fields were measured by directly detecting each field and comparing their photo-currents for correlation \cite{ou92PRL}, but this method fails when atomic states are involved because we cannot directly measure the atomic states. Even though the atomic state can be read out by a reading beam, which converts the atomic spin wave $\hat S_A$ into an anti-Stokes field for measurement \cite{kuz03}, it is not easy to find matching local oscillators for both Stokes and anti-Stokes to measure the quadrature-phase amplitudes. So, the quantum correlation between the quadrature-phase amplitudes of the atoms and the
Stokes field was never measured. However, it was recently proposed and demonstrated \cite{shaked18,li19,li20} that a high-gain parametric amplifier can assist in measuring the quantum correlation between two fields that are input as signal and idler to the amplifier and the detection of only one output field, say, the signal, is enough to find the information. The Raman amplifier, which in essence is a parametric amplifier, can serve this purpose here without measuring atomic fluctuations. In fact, if we look at Eq.(\ref{R}) and set $G \gg 1$ so that $\lambda\approx 1$, we have
\begin{eqnarray}\label{R2}
R(\lambda=1) =  \frac{1}{2}\langle \Delta^2(\hat X_A^{(in)} + \hat X_S^{(in)})\rangle = \frac{1}{2} \Delta^2 X_+,
\end{eqnarray}
which gives the quantum correlation $\Delta^2 X_+$ between the atomic spin wave ($\hat X_A^{(in)}$) and the Stokes light field $(\hat X_S^{(in)})$ compared to the uncorrelated value of 2 for vacuum.

\begin{figure}[t]
\centering
\includegraphics[width=1.0\columnwidth]{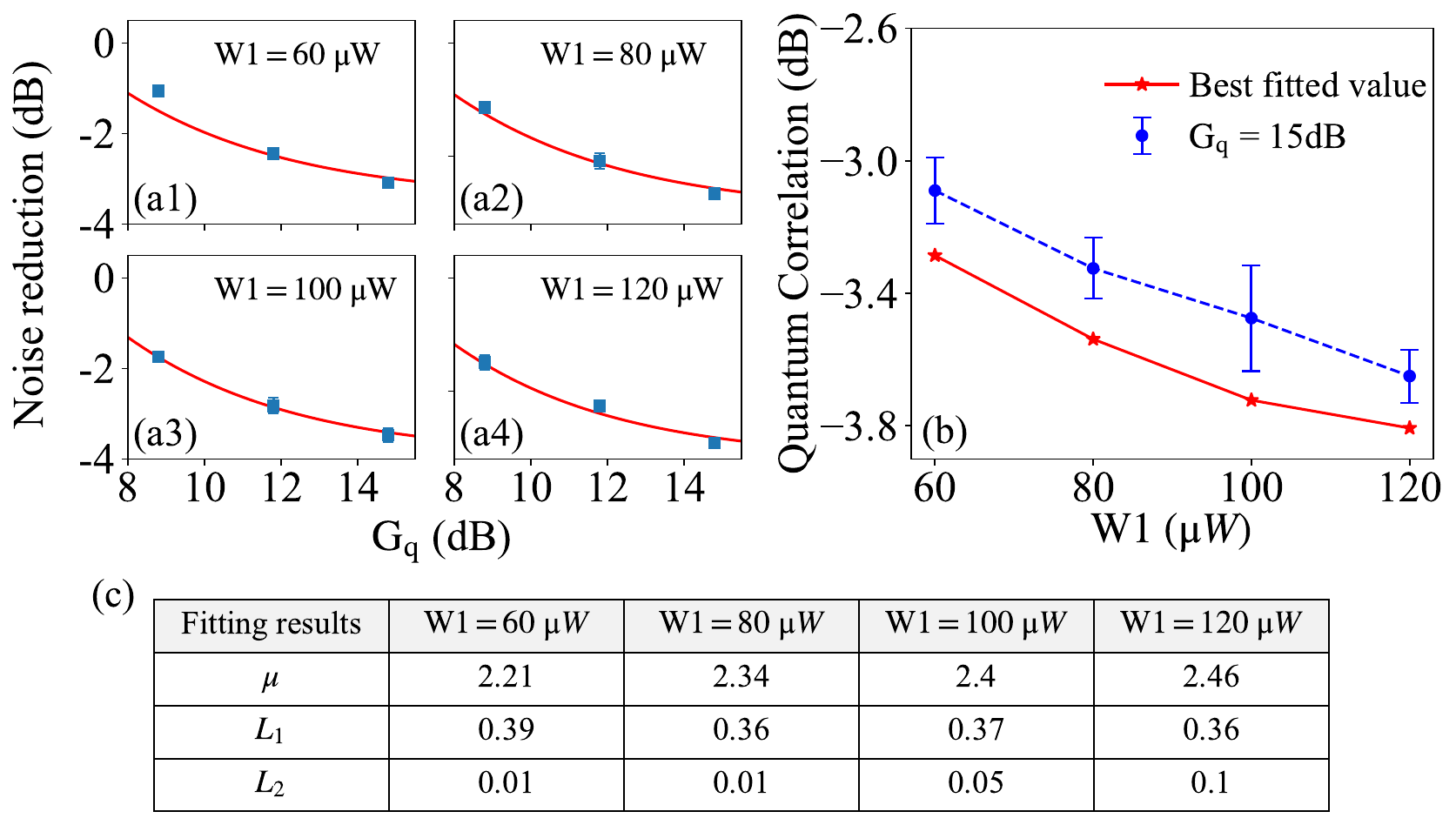}
	\caption{Noise level reduction as a function of the gains of RA2 for different degrees of correlation between atom and light. (a1-a4) are fitted via Eq.(\ref{R3}) to get the quantum correlation coefficients at different powers of W1. (b) Solid red points are the quantum correlation $\Delta^2 X_+$ in log-scale with 0 dB corresponding to vacuum, evaluated using best fit parameters obtained in (a). (c) The fitting values of $\mu$, the Stokes field loss $L_1$, and the atomic spin wave loss $L_2$ obtained from Eq.(\ref{R3}) at different powers of W1. } 
	\label{Fig4}
\end{figure}

So we now include RA2 as a part of the measurement tool to show the quantum correlation between the Stokes field and the atomic spin wave generated from RA1, as shown in the conceptual schematic in Fig.\ref{Fig1}(b). To make this measurement, according to Eq.(\ref{R2}), we need to operate RA2 at high gain. In fact, we see that the brown trace in Fig.\ref{Fig3}(b) is close to this case with a quantum gain of 15 dB or $G_q= 32 \rightarrow \lambda = \sqrt{G^2-1}/G = \sqrt{(G_q-1)/(G_q+1)}= 0.97$. However, this is still not 1. So, we next use all the data at different $G_q$ of RA2 to extract the limiting case of $\lambda\rightarrow 1$. The other traces with smaller RA2 gains have lower noise reduction values than the brown trace due to smaller $\lambda$. In order to have a better estimate of the quantum correlation $\Delta^2 X_+$ between atoms and light from RA1 and to make full use of all noise reduction data collected,  we re-plot the data in Fig.\ref{Fig3}(b) but using the quantum gain $G_q$ of RA2 as a variable, as shown in Fig.\ref{Fig4}(a1-a4) for different W1. We can use a simple model given in Ref.\cite{li20} for both RA1 and RA2 by assuming that RA1 is also described by a parametric amplifier but with amplitude gains of $\mu$ and $\nu=\sqrt{\mu^2-1}$. We also introduce losses for Stokes ($L_1$) and the atomic spin wave ($L_2$) to account for imperfect mode match and propagation losses, as shown in Fig.\ref{Fig1}(b).  With proper phase of the inputs for minimum noise, it is straightforward to evaluate Eq.(\ref{R}) for the scheme in Fig.\ref{Fig1}(b) as
\begin{eqnarray}\label{R3}
R &=& \mu^2+\nu^2 -2\nu^2\frac{L_2+L_1\lambda^2}{1+\lambda^2} 
\cr &&- \frac{4\lambda\mu\nu\sqrt{(1-L_1)(1-L_2)}}{1+\lambda^2},
\end{eqnarray}
with $\lambda \equiv g/G = \sqrt{(G_q-1)/G_q+1)}$. 
We then try to find $\mu,\nu, L_1, L_2$ by fitting the theory above to the data in Fig.\ref{Fig4}(a1-a4). The results are shown as solid red traces.  The best fitted values of $\mu,\nu$ and $L_1,L_2$ for each W1 (see Fig.\ref{Fig4}(c)) are then used to evaluate the quantum correlation $\Delta^2 X_+ = 2R(\lambda=1)$ from Eq.(\ref{R3}) \cite{li20}, which is plotted as a function of W1 as shown in Fig.\ref{Fig4}(b) in log-scale with 0 dB corresponding to the vacuum level or $\Delta^2 X_+^v = 2$. The trend aligns well with the predictions that increasing the W1 power strengthens the quantum correlation between the Stokes field and the atomic spin waves from RA1. For comparison, we also evaluate the quantum correlation using the brown data in Fig.\ref{Fig3}. The results are plotted as the blue dashed trace. As expected, the values are close to but a bit higher than the red ones because $\lambda = 0.97\lessapprox 1$. 

As discussed above, the amount of noise reduction highly depends on the effective losses of the Stokes field (S1) between RA1 and RA2. But we never mention the losses of the output (S2), which is measured as 2.8 dB overall. In fact, it does not matter here because the observed quantum noise from RA2 is more than 20 times the vacuum noise, making the vacuum noise leaked from loss irrelevant. This is the advantage of using the parametric amplifier as a part of the measurement \cite{li19,li20}.
%According to Eq.(\ref{R}), the dependence of the noise reduction on the quantum gain of RA2 roughly follows ${\rm var}/(1+\lambda^2) = {\rm var} G^2/(2G^2-1) = {\rm var } (G_q+1)/2G_q$, where ${\rm var}\equiv \langle \Delta^2( \hat X_a^{(in)} + \lambda \hat X_S^{(in)})\rangle$. Thus, using the experimental noise reduction and $G_q$, we solve for theoretical var, as shown by the red solid lines in Fig.\ref{Fig4}(a1)-(a4). For the best fit, a coefficient was applied to $G_q$, reflecting the actual $G_q$ used. The var measured at $G_q=12~\mathrm{dB}$ under different powers of W1 is shown in Fig.\ref{Fig4}(b). The trend is consistent with previous predictions that as power of W1 increases, the quantum correlation between the Stokes field and atomic spin waves generated by RA1 becomes stronger.

% According to Eq.(\ref{R}), the dependence on the quantum gain ($G_q=2G^2-1$) of RA2 should roughly follow ${\rm var}/(1+\lambda^2) = {\rm var} G^2/(2G^2-1) = {\rm var } (G_q+1)/2G_q$ as depicted in the solid curves with ${\rm var}\equiv \langle \Delta^2( \hat X_a^{(in)} + \lambda \hat X_S^{(in)})\rangle$ adjusted for the best fit. 

\begin{figure}[t]
\centering
\includegraphics[width=0.6\columnwidth]{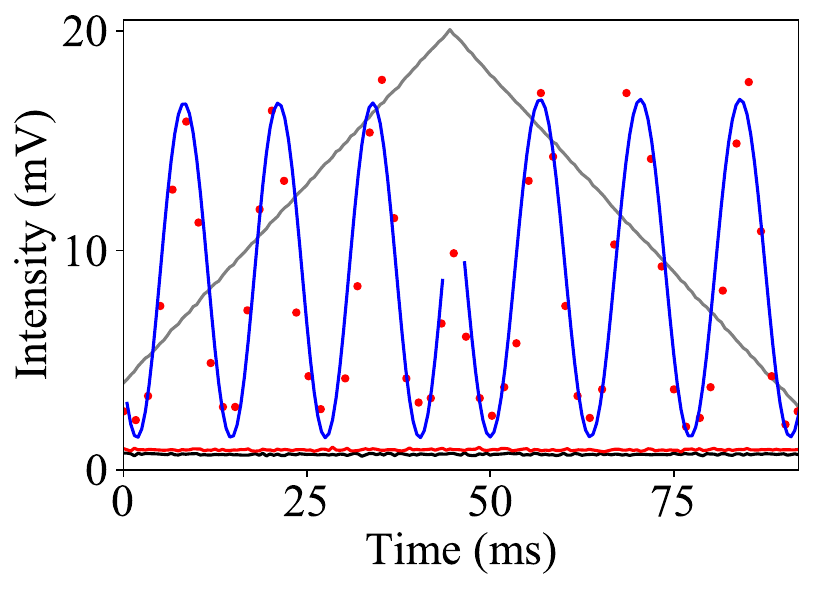}
	\caption{Interference fringes (red dots with blue trace fit to sine-function) as a function of phase scan (gray), measured at the output of RA2. Red trace: injected Stokes level at input of RA1; Black trace: background level.} 
	\label{Fig5}
\end{figure}

 Next, notice that although we demonstrated that amplifier's noise is reduced here by exploiting quantum correlation, the method cannot be used to amplify an arbitrary (unknown) signal since the input is required to have correlation with the atoms. On the other hand, our scheme can be used as a quantum sensor of phases as described next. If the two Raman amplifiers of RA1 and RA2 are treated as one system, as shown in Fig.\ref{Fig1}(c), they form an SU(1,1) interferometer, which uses parametric amplifiers to replace traditional linear beam splitters for wave splitting and addition for interference \cite{ou20}. Here, the parametric amplifiers are the Raman amplifiers of RA1 and RA2. To see the interference fringes, we inject a coherent beam at the Stokes input of RA1 and detect the output Stokes field with a power detector \cite{chen15}. Figure \ref{Fig5} shows the interference fringes as the phase of the back-injected Stokes field S1 is scanned.

In conclusion, by preparing the internal atomic state of a Raman amplifier in a quantum-correlated state with the input Stokes field through another Raman process, we are able to achieve close to 4 dB reduction of the amplifier's output noise. This approach may have potential for quantum-enhanced measurements in Raman spectroscopy\cite{mic19,cas21}. Additionally, the scheme with two Raman amplifiers in tandem can be used to measure the quantum correlation generated by the first Raman process by placing the second Raman amplifier in high-gain mode. This provides the first measurement of the atom-light field correlation in the Raman process. With a seed field injected into the Stokes field of the system, an SU(1,1) interferometer can be formed that is sensitive to both optical and atomic phases.
% \begin{figure}[t]
% \centering
% \includegraphics[width=0.6\columnwidth]{Figure/Fig5.pdf}
% % \includegraphics[width=4.3cm]{Figure/Fig-gain.jpg}
% 	\caption{Interference fringes (red dots with blue trace fit to sine-function) as a function of phase scan (gray), measured at the output of RA2. Red trace: injected Stokes level at input of RA1; Black trace: background level.} 
% 	\label{Fig5}
% \end{figure}

% \begin{figure}[h]
% \centering
% \includegraphics[width=1.0\columnwidth]{Figure/Fig3_v2.pdf}
% % \includegraphics[width=4.3cm]{Figure/Fig-gain.jpg}
% 	\caption{} 
% 	\label{Fig3_V2}
% \end{figure}
% % \begin{figure}[ht]
% % \centering
% % \includegraphics[width=1.0\columnwidth]{Figure/Fig4.pdf}
% % % \includegraphics[width=4.3cm]{Figure/Fig-gain.jpg}
% % 	\caption{} 
% % 	\label{Fig4}
% % \end{figure}

% \vskip 0.2in

\begin{acknowledgments}
\textit{\bf Acknowledgments.} The work is supported by City University of Hong Kong (Project No.9610522) and the General Research Fund from Hong Kong Research Grants Council (No.11315822).

\end{acknowledgments}
\vskip 0.2in
\noindent {\bf AUTHOR DECLARATIONS}

The authors have no conflicts of interest to disclose.

%\section*{References}

\bibliography{main}% Produces the bibliography via BibTeX.

% \bibitem{polzik} K. Hammerer, A. S. Sørensen, E. S. Polzik, “Quantum interface between light and atomic ensembles,” Rev. Mod. Phys. 82, 1041 (2010).

% \bibitem{sha18} Y. Shaked, Y. Michael, R. Z. Vered, L. Bello, M. Rosenbluh, and A. Pe’er, “Lifting the bandwidth limit of optical homodyne measurement with broadband parametric amplification,” Nat. Commun. 9, 609 (2018).

% \bibitem{ou92} Z. Y. Ou, S. F. Pereira, H. J. Kimble, and K. C. Peng, “Realization of the Einstein-Podolsky-Rosen paradox for continuous variables,” Phys. Rev. Lett. 68, 3663 (1" 992). https://doi.org/10.1103/PhysRevLett.68.3663

\end{document}